\documentclass{svjour3}
\usepackage{amsmath}
\usepackage{graphicx}
\usepackage[numbers,round,sort&compress]{natbib}
\renewcommand{\cite}[1]{\citep{#1}} 
\newcommand{\argmax}{\operatornamewithlimits{argmax}}
\newcommand{\mmax}{\operatornamewithlimits{max}}
\begin{document}
\bibliographystyle{chicago}

\title{The pattern and process of gene family evolution}
\author{Gergely J Sz\"oll\H{o}si and Vincent Daubin}
\institute{Gergely J Sz\"oll\H{o}si \at
UMR CNRS 5558 - LBBE
"Biométrie et Biologie Évolutive"
UCB Lyon 1 - Bât. Grégor Mendel
43 bd du 11 novembre 1918
69622 VILLEURBANNE cedex
\email{Gergely-Janos.Szollosi@univ-lyon1.fr}
\and
Vincent Daubin \at
UMR CNRS 5558 - LBBE
"Biométrie et Biologie Évolutive"
UCB Lyon 1 - Bât. Grégor Mendel
43 bd du 11 novembre 1918
69622 VILLEURBANNE cedex
\email{Vincent.Daubin@univ-lyon1.fr}
}

\maketitle
\begin{abstract}
Large scale databases are available that contain homologous gene families constructed from hundreds of complete genome sequences from across the three domains of Life. Here we discuss approches of increasing complexity aimed at extracting information on the pattern and process of gene family evolution from such datasets. In particular, we consider models that invoke processes of gene birth (duplication and transfer) and death (loss) to explain the evolution of gene families. 

First, we review birth-and-death models of family size evolution and their implications in light of the universal features of family size distribution observed across different species and the three domains of life. Subsequently, we proceed to recent developments on models capable of more completely considering information in the sequences of homologous gene families through the probabilistic reconciliation of the phylogenetic histories of individual genes with the phylogenetic history of the genomes in which they have resided.

To illustrate the methods and results presented, we use data from the HOGENOM database, demonstrating that the distribution of homologous gene family sizes in the genomes of the Eukaryota, Archaea and Bacteria exhibit remarkably similar shapes. We shown that these distributions are best described by models of gene family size evolution where for individual genes the death (loss) rate is larger than the birth (duplication and transfer) rate, but new families are continually supplied to the genome by a process of origination. Finally, we use probabilistic reconciliation methods to take into consideration additional information from gene phylogenies, and find that, for prokaryotes, the majority of birth events are the result of transfer. 
 
\end{abstract} 

\section{ Introduction}

The strongest evidence for the universal ancestry of all life on Earth comes from two sources i) the shared molecular characters essential to the functioning of the cell, such as fundamental biological polymers, core metabolism and the nearly universal genetic code; ii) sequence similarity between functionally related proteins in the Bacteria, Archaea and Eukaryota\cite{Crick:1968dz,Theobald:2010cr}. However, the majority of functionally related genes, similar to other phylogenetic characters, exhibit a more restricted distribution and consequently taken separately, can only provide phylogenetic information on finer scales. Nonetheless, considered together the ensemble of related sequences carry a comprehensive record of the evolutionary history and mechanisms that have generated them\cite{Boussau:2010cr}. Sequence similarity on these finer scales has been used to construct large scale databases of putative sets of sequences of common ancestry, in particular homologous proteins and protein domains. At present such databases constructed from hundreds of complete genome sequences from across the three domains of Life are available. Here we discuss methods capable of extracting information on the pattern and process of genome evolution from large scale datasets composed of \emph{homologous gene families}. 

\section{ Birth-and-death processes and the shape of the protein universe}

The majority of bacterial, archaeal and eukaryotic genes belong to homologous families \cite{Koonin:2008kx} which together contain a potential treasure trove of information on the pattern and process of descent of these genes, and the genomes in which they reside. A qualitative examination of the number of family members in genomes and the phylogenetic distribution of the families reveals two important patterns: i) the distribution of the majority of homologous gene families is not universal, but phylogenetically limited and ii) many families contain multiple members from the same genomes, while at the same time, being characterized by a patchy distribution. These observations imply that i) some process of \emph{gene origination} must exist that result in the ongoing generation of sequences sufficiently different to be seen as a novel gene family and ii) processes of \emph{gene birth} capable of creating new genes with recognizable homology from existing ones must also exists in parallel with processes of \emph{gene death} leading to the loss of existing genes. 

Considering the latter case first, several molecular mechanisms are known to be involved in the creation of new gene structures in a genome. Among eukaryotes, a range of mechanisms are know to be capable of producing gene-sized duplications of genetic material\footnote{ These mechanisms include exon shuffling, reverse transcription of expressed genes and the action of mobile elements for reviews see \cite{Long:2003vn,Lynch:2007bh}.}. In the case of prokaryotes, mechanisms for duplication are less well understood and horizontally transfered genes are believed to be an important, perhaps dominant, source of new gene structures entering the genome\footnote{Transfer of DNA into the prokaryotic cell can occur primarily by three means: (i) transduction by viruses, (ii) conjugation by plasmids, and (iii) natural genetic transformation the ability of some bacteria to take up DNA fragments released by another cell. For details see \cite{Gogarten:2005qf}.}\cite{Lerat:2005ys}. While we expect duplication to produce gene copies with recognizable homology, whether transfer is seen as gene origination or gene birth in the context of a particular genome depends on the presence of recognizable homologs. In contrast to duplication and transfer, the loss of genes are thought to most frequently result from a cascade of small deletion events with small or no fitness effect, which follow the initial inactivation of a gene (the emergence of a pseudogene). As in the case of pseudogenization, molecular mechanisms can generate new gene structures or lead to the loss of existing ones in the genomes of individual cells, the fate of these genomic changes, whether they will fix or be lost in the population, will be determined by their selective effects and population genetic paramaters, such as effective population size.

On the broadest scale the strength of genetic drift has been hypothesized to be a dominant factor influencing genome size across all three domains of life \cite{Lynch:2003ff}. As we will see in the following section, the pattern of the distribution of homologous gene family sizes in and among genomes can, to a large extent, also be described in terms of essentially neutral stochastic birth-and-death processes. Birth (duplication and transfer) and death (loss) in the context of these models correspond to the addition and removal of genes to homologous gene families over evolutionary time-scales that are long compared to the mutational and population genetic time-scales.

The question of mechanisms responsible for the origination of gene families is not well understood. A significant fraction of genes, in genomes from all three domains of life appears to be of very recent origin in so far as they are restricted to a particular genome and possess no known homologs. By some counts, such orphan genes constitute, e.g.\ one third of the genes in the human genome \cite{Lynch:2007bh} and $14\%$ in a survey of 60 bacterial genomes \cite{Siew:2003zr}. While there are signs that a large fraction of orphan genes in prokaryotic genomes may have viral origin \cite{Daubin:2004ff}, our understanding of where these genes come from, more generally what the dominant processes of gene origination are, remain largely unresolved fundamental questions. Nevertheless, as we show below using birth-and-death processes as models, the continuous presence and significance of origination during the course of genome evolution is readily apparent from the record it has in the pattern of gene homologous family sizes, i.e., in the shape of the protein universe. 

\subsection{The distribution of homologous gene family sizes}

\begin{figure}
\centerline{\includegraphics[width=1.\textwidth]{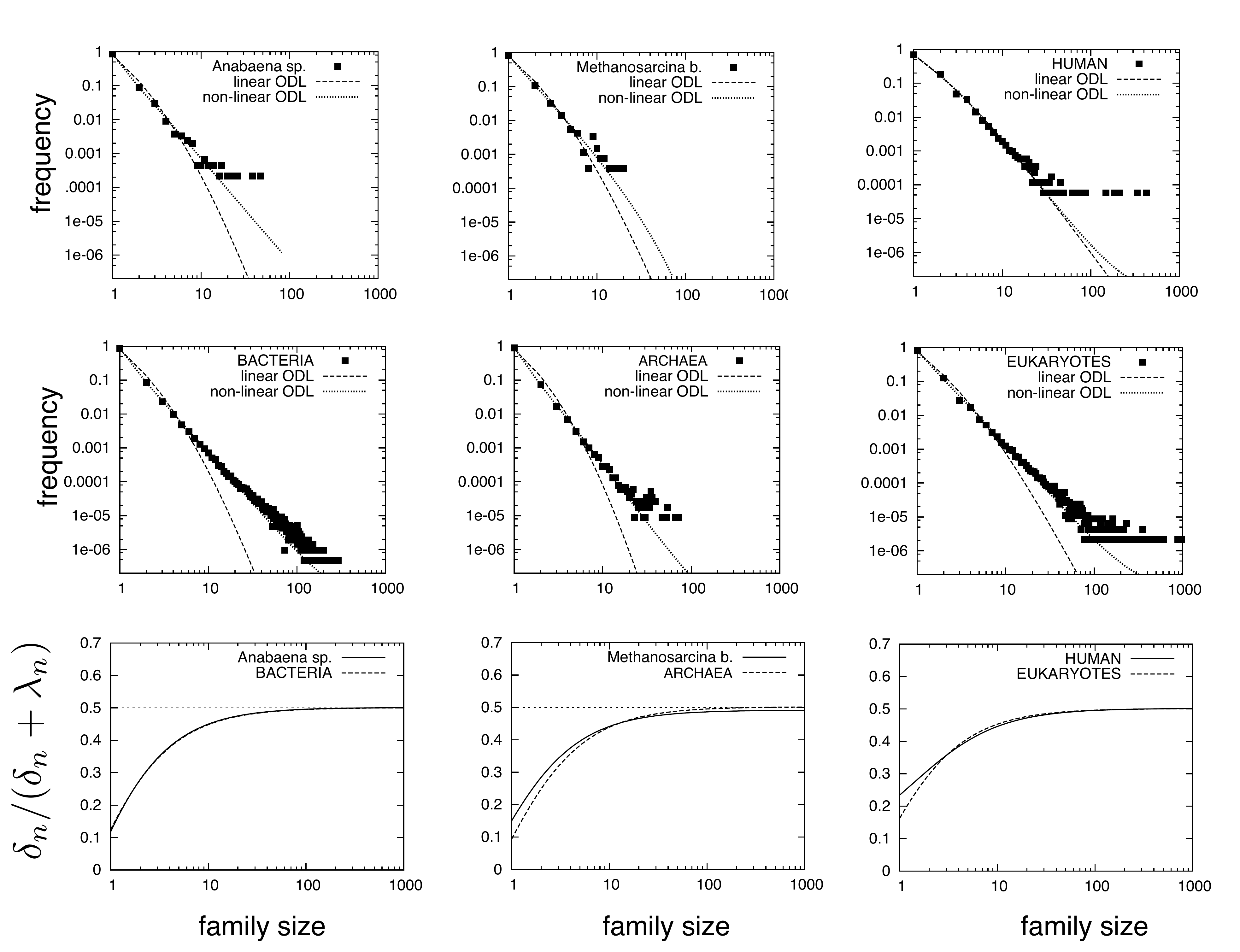}}
\caption{ \textbf{Distribution of homologous gene family sizes across the three domains.}
The distribution of homologous gene family sizes was derived from the version 5 of the HOGENOM database \cite{Penel:2009ly}. The results for the three domain data for the complete genomes of $820$ Bacteria, $62$ Archaea and $64$ Eukaryotes, and correspond to the average of the frequencies of family sizes across species in the domain. Dashed lines indicate fits with different origination duplication and loss (ODL) models. The linear model corresponds to the model of Reed et al., the non-linear is that proposed by Karev et al., see text for details. The bottom row presents the relative rate of duplication as a function of family size corresponding to the fits of the non-linear model of equation \ref{karev_eq} in the two rows above it. 
}
\label{BD_plots}
\end{figure}

The frequency distribution of gene family sizes in the complete genomes of organisms from all three domains exhibit remarkably similar shapes with characteristic long, slowly decaying tails \cite{Huynen:1998ve,Qian:2001bh,Karev:2002dq}. These distribution all have a power-law shape, for large family size $n$ the frequency of families $f(n)$ falls off as $f(n) \propto n^{\gamma}$ with some $\gamma<0$. This power-law shape is apparent in the log-log plots of figure \ref{BD_plots} and corresponds to an excess of large and very large families compared to what would be expected based on the size of the average gene family. Even more remarkable is the similarity of the family size distributions between species from a single domain (columns in figure \ref{BD_plots}), and even between domains (rows in figure \ref{BD_plots}). This similarity implies that the process that have generated these distributions may share universal features across species and across the three domains. Here we focus on the information that can be inferred under the assumption that particular forms of birth-and-death processes have shaped these distributions and will not consider potential connections with power-law scaling in functional genome content \cite{Molina:2009bs} or homology networks and their connection to other biological networks with similar characteristics \cite{Koonin:2006ij}. 

\subsection{Interpreting the pattern of gene family sizes}

\label{interpretpattern}
Huynen and van Nimwegen were the first to describe and interpret a widespread pattern of a slowly decaying asymptotic power-law in the distribution of homologous gene family sizes. They examined a diverse set of genomes spanning the Bacteria, Archaea, Eukaryota and viruses \cite{Huynen:1998ve}. They found that a simple, but relatively abstract, stochastic birth-and-death process, one where the duplication and loss events are correlated within a family, produces power-law distributions (for details see below). They found the exponent $\gamma$ to be between $-2$ and $-4$ in their studies. In fact, a value consistent with these results of $\gamma$ between $-2$ and $-3$ has been observed in all subsequent studies and can easily be read off from figure \ref{BD_plots}. In the context of Huynen and van Nimwegen's model this indicates that the origination rate (in general a combination of gain resulting from transfer, and the birth of new families with no homologous in other genomes) that is required to compensate for the stochastic loss of families must be significant.

Subsequent work has shown that for models where the birth and death of genes in a gene family are considered independent the asymptotic decay of the distribution of gene family sizes can also become a power-law. Albeit such behavior is only exhibited by a certain, specific subclass of origination-duplication-loss type birth-and-death models. As demonstrated by \cite{Karev:2002dq}, this is the case for non-linear models (see below) in which the death rate approaches the birth rate for large families, but is considerably greater than the birth rate for small families (see bottom row of figure \ref{BD_plots}). Karev et al.\ have been able to accurately reproduce the distributions of gene (and domain) family sizes for a range of analysed genomes. The origination rates necessary to fit empirical family size distributions were found to be relatively high, and comparable, at least in small prokaryotic genomes, to the overall intra-genomic duplication rate. This has been interpreted as support for the key role of horizontal gene transfer in these genomes \cite{Novozhilov:2006qf,Karev:2002dq,Koonin:2002rt}.

At about the same time as the work of Karev and colleagues appeared, Reed et al.\ demonstrated \cite{Reed:2004bh} that a very simple birth-and-death process can also exhibit an asymptotic power-law. They considered a model where the birth and death of genes are independent of each other and of family size, and origination occurs randomly with a uniform rate (see below), and found asymptotic power-law behavior under the condition that the rate of birth (duplication) is larger than the rate of death (loss). In figure \ref{BD_plots}, we show comparisons of the fits of the linear model of Reed et al.\ and the non-linear model of Karev et al.\ to gene family size distributions for the three domains. We can see that despite its relative simplicity, considering data from individual species (top row of figure \ref{BD_plots}), the linear model (described by three parameters) provides comparable quality fits as the model of Karev et al.\ (described by five parameters). If we consider, however, the fits to distributions averaged over the three domains\footnote{As the functions being fit are discrete probability distributions, one can easily calculate the probability of the observed empirical distribution given values of the model parameters, and subsequently perform fitting by maximizing the likelihood of the model parameters. For the case of the averaged distributions this method of fitting using likelihood allows a clear interpretation of the fit to the averaged distributions, as corresponding to the hypothesis of a birth-and-death process with identical parameter values across all species in the domain having generated the observed distribution. }, we can observe that the non-linear mode clearly provides a better fit (second row of figure \ref{BD_plots}). Perhaps more conclusively the parameter values obtained in the case of the linear model, corresponding to a birth to death ratio of between roughly 2 and 5 ($\delta/\lambda = 4.9$ for the Human dataset with the best apparent fit) is qualitatively at odds with empirical estimates of the recent duplication and loss rates in eukaryotic genomes, which unanimously indicate a value much smaller than one (see table 8.1 in \cite{Lynch:2007bh}). 

\begin{figure}
\centerline{\includegraphics[width=0.6\textwidth]{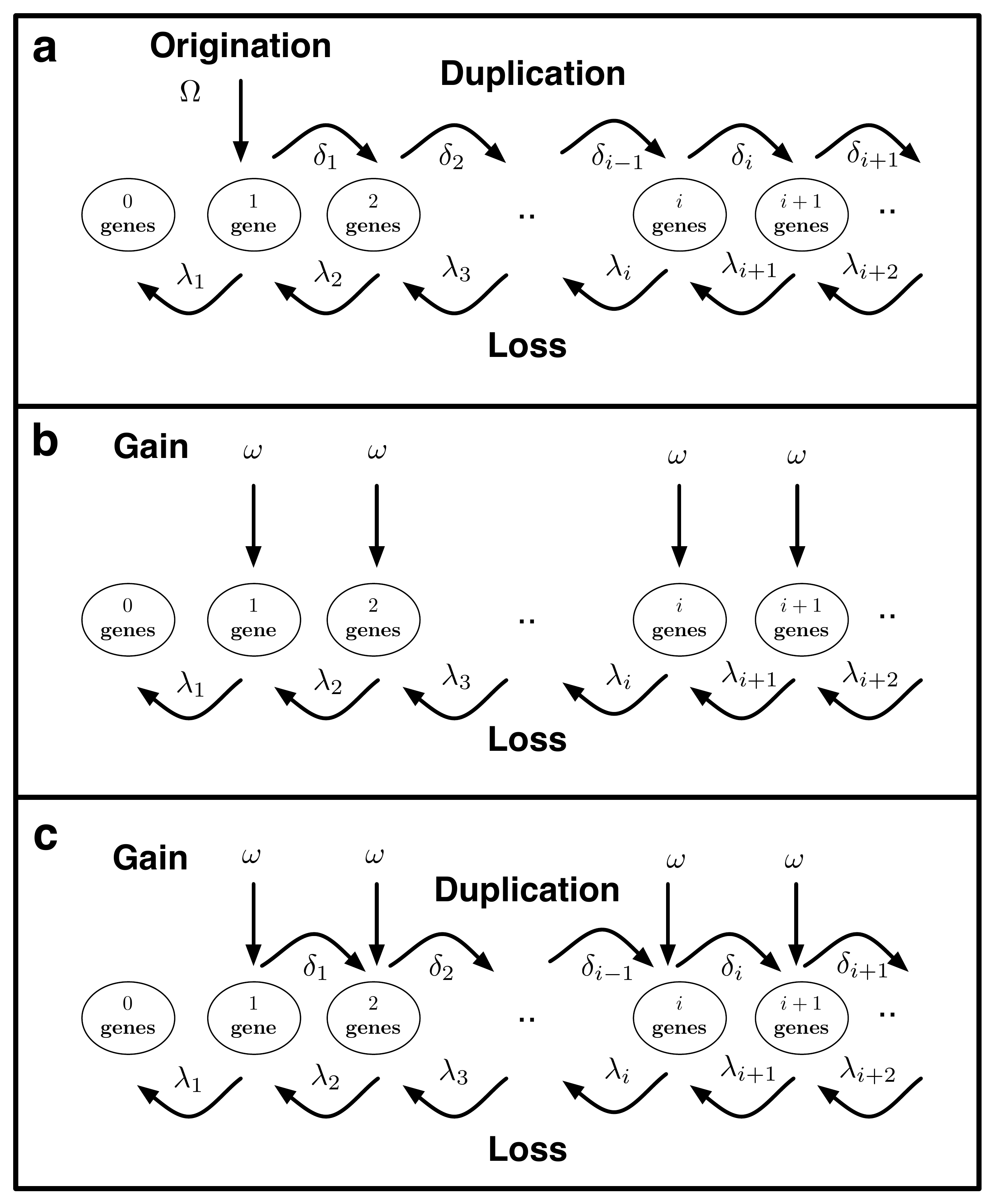}}
\caption{
\textbf{Birth-and-death models of homologous gene family evolution}
A birth-and-death process is a stochastic process in which transitions between states labeled by integers (representing the number of individuals, cells, lineages, etc.) are only allowed to neighboring states. A jump to the right constitutes birth, whereas a jump to the left is a death. In the context of birth-and-death processes that model the evolution of homologous gene families, the number of representatives a homologous gene family has in a given gene corresponds to the model state. Birth represents the addition of gene to a family in genom as a result of: i) origination of a new family with a single member, duplication of an existing gene, or gain of a gene by means of horizontal transfer of a gene from the same family from a different genome. The three models pictured above have been used in different contexts to model observed patterns of gene family size: \textbf{a}) the stationary distribution of non-linear origination-duplication-loss type models are able to reproduce the general shape and in particular the power-law like tail of the distribution of homologous gene family sizes (cf.\ section \ref{interpretpattern} and \cite{Karev:2002dq}), while transient distributions of linear origination-duplication-loss can be used to construct models of gene family size evolution along a phylogeny, modeling the ``inparalog'', i.e., vertically evolving component of the size family distribution \cite{Csuros:2009zr}; \textbf{b}) and \textbf{c}) linear gain-loss and gain-duplication-loss type models are used to model the non-vertically evolving, so called ``xenolog'', component of the family size distribution along a branch of a phylogenetic tree. 
}
\label{BD}
\end{figure}

\subsection{The theory of birth-and-death processes}

Historically, the biological application of birth-and-death processes, starting with the seminal work of Yule \cite{Yule:1925ys} in the 1920s, and continuing in the following decades \cite{Feller:1939,Kendall:1948yq,Bartholomay:1958kx,Takacs:1962tg}, was the construction of stochastic models that can furnish a means for interpreting random fluctuations in the population size with time. The application of birth-and-death process to sizes of gene families is more recent. The realization that the sizes of gene families can be compared with the aim of better understanding adaptive evolutionary processes and organismal phylogeny began with the work of Hughes and Nei \cite{Ota:1994uq,Nei:1997fk} and others \cite{Yanai:2000vn} in the context of the debate on whether differences in the copy number of major histocompatibility complex genes across species has evolved due to adaptive or stochastic forces. As described above recent work has focused on explaining the distribution of the number of genes in homologous gene families in genomes as the result of stochastic birth-and-death processes (see also Chapter 3 of \cite{Lynch:2007bh}). 

A birth-and-death process is a stochastic process in which transitions between states labeled by integers (representing the number of individuals, cells, lineages, etc.) are only allowed to neighboring states (see figure \ref{BD}). An increase by one of the number of individual (or genes in a gene family) constitutes birth, whereas decrease by one is a death. More formally the dynamics of a population (of individuals, or of genes in a gene family) is represented by a Markov process, i.e., the state of the population at time $t$ is described by the value of a random variable described by the Markov property (for an accesible review see \cite{Novozhilov:2006qf}). In general, for each state the probability of both birth, a transition from state $n$ to $n+1$ and of death, a transition from state $n$ to $n-1$ is described by a rate birth rate $\delta_n$ and a death rate $\lambda_n$. A third elementary process besides birth and death that is relevant in the context of gene family size evolution is origination. As described above, not all gene families are of the same age, consequently to model the process of origination of new families, families with a single gene relavant to originate at some rate constant $\Omega$ as shown in figure \ref{BD}. Considering a similar rate of influx into each state can be regarded as a model of horizontal gene transfer (HGT) cf.\ figure \ref{BD}. 

The simplest type of birth-and-death processes with biological relevance are linear birth-and-death processes. Linear birth-and-death processes are described by a single birth rate $\delta$ and a single death rate $\lambda$ from which the state-wise rates can be derived by the following first order rate law:
\begin{equation}
\label{hughes_eq}
\delta_n= \delta n \quad \mathrm{and} \quad \quad \lambda_n=\lambda n.
\end{equation}
In other words, a gene (individual) in a gene family (population) gives birth to a new gene at a rate $\delta$ and undergoes death at a rate $\lambda$, independent of the size of the gene family. The stationary distribution of a linear birth-and-death process with origination --with some rate $\Omega$-- can be shown to be i) a stretched exponential if $\delta\leq \lambda$, i.e., the birth rate is smaller than the death rate or ii) exhibits an asymptotic power-law behavior with exponent $\gamma=(\Omega/(\delta-\lambda)+1)$ \cite{Hughes:2005mi} if $\delta> \lambda$. The transient distribution can be analytically expressed for the linear version of all three processes shown in figure \ref{BD} . These distributions are important in deriving the probability of observing a particular pattern of family sizes at the leaves of a phylogeny, as well as in estimating branch-wise duplication, transfer and loss parameters from a forest of gene trees that have been mapped using a series of duplication transfer and loss events to the branches of a species phylogeny (see section \ref{phyloBD}). 
 
A succession of more complex nonlinear models can be constructed, the simplest proposed \cite{Karev:2002dq} being a model with a family size dependent duplication and loss rate parametrized by a pair of constants $a$ and $b$:
\begin{equation}
\label{karev_eq}
\delta_n= \delta(n) n=\left( \frac{\delta' (n+a)}{n} \right) n \quad \mathrm{and} \quad \quad \lambda_n=\lambda(n) n =\left( \frac{\lambda' (n+b)}{n} \right) n,
\end{equation}
where we have not simplified by $n$ to emphasize the relationship with the linear model above. For this class of models, asymptotic power-laws are obtained only if $\delta'<\lambda'$ \cite{Karev:2002dq}, i.e., the birth rate is smaller than the death rate\footnote{It is important to note that the linear origination-duplication-loss type model of Reed et al.\ \cite{Reed:2004bh} differ from those of Karev et al.\ \cite{Karev:2002dq} in details related to how origination is considered and in how the space of possible states (family sizes) and hence the stationary state is defined. While Hughes and Reed consider gene families to originate at a constant rate and consider family size to be unbounded, Karev et al.\ assume that family sizes are bounded and consider reflecting boundary conditions. }. Discrete time models that are closely related to the continuous time models considered by Karev et al. were presented by W\'ojtowicz and Tiurjn\cite{Wojtowicz:2007ys}.

A different more abstract type of birth-and-death process was historically the first to be proposed to model the distribution of gene family sizes \cite{Huynen:1998ve}. Similarly to the above model, a gene family is founded by a single ancestor, and the size of the family may change as a result of duplications and losses (birth and death). However, in contrast to the birth-and-death models considered so far, duplications and losses are considered to act ``coherently'' on genes within one gene family. That is, if a certain gene is likely to duplicate (be lost), then all genes of its family are likely to duplicate (be lost). More formally, denoting the size of a gene family at time $t$, by $n_t$ 
\begin{equation}
\label{huynen_eq}
n_{t} = \alpha_t n_{t-1},
\end{equation}
where $\alpha_t$ is a random multiplication factor, giving the instantaneous ratio of birth to death, that is drawn independently at each time step from some distribution $P(\alpha)$. The distribution of gene family sizes that is the result of many such processes can be shown to have a power-law distribution, provided the further important condition that some form of origination be present is met. The exponent of the power-law asymptotic followed by the family size distribution will in this case be independent of the exact nature of origination (independent e.g.\ of whether one considers reflecting boundary conditions or random influx) and is given by $\gamma=-(1-\mu_\alpha/\sigma^2_\alpha)$, where $\mu_\alpha = \langle \log(\alpha) \rangle$ is the mean of the logarithm of the random variable $\alpha$ and $\sigma^2_\alpha= \langle \log^2(\alpha) \rangle - \mu_\alpha$ is its variance \cite{Huynen:1998ve}. Interestingly, this implies that birth-and-death models with coherent noise (also called multiplicative noise) produce a power-law asymptotic irregardless of whether the birth rate is smaller or larger than the death rate. The value of the exponent, however, can give an indication of their relative values. The reason being that since $\sigma^2_\alpha$ is positive, $\gamma<-1$ implies $\mu_\alpha=\langle \log(\alpha) \rangle<0$, which can be shown to be equivalent to the geometric mean of $\alpha$, i.e., the instantaneous ratio of birth to death, being smaller than unity. 

\subsection{Birth-and-death along a species phylogeny}
\label{phyloBD}
So far, we have only considered the distribution of homologous gene family sizes in genomes of individual species and the average of such distribution across domains. The distributions of gene family sizes between species are, however, not independent, but rather reflect correlated histoires related by common descent along a species phylogeny. The phylogenetic profile of a gene family, consisting of the number of homologs within the same family in each genome, encodes this information. Such phylogenetic profiles can be informative even though they neglect a large part of the information present in gene sequences. Nonetheless, profile data sets have been used both to construct organismal phylogenies \cite{Fitz-Gibbon:1999ly,Snel:1999ve,Wolf:2002bh,Deeds:2005dq,Lienau:2006nx} and reconstruct ancestral gene content \cite{Mirkin:2003dq}. These methods have, however, proved sensitive to methods of homology inference and have relatively poor performance as methods of phylogenetic analysis. This can been explained, in the case of prokaryotes by high levels of homoplasy\footnote{Homoplasy ( also called convergent evolution) describes the acquisition of the same biological trait (in this case, genes from the same family) in unrelated lineages.} resulting from both horizontal gene transfer, but also and extensive parallel loss of gene families in certain bacteria genomes \cite{Hughes:2005mi}.

The primary advantage of the above attempts at reconstructing phylogeny is their relative ease of implementation and computational tractability on large datasets derived from complete genomes. They, however, suffer two major shortcomings: i) they lack an explicit model of evolution and consequently provide at best indirect information on processes, and ii) they disregard a great deal of phylogenetically relevant information present in homologous sequences, by considering only presence-absence, or at most the gene copy number in genomes. 

The first of these shortcomings can be overcome by considering phylogenetic profiles as observations at the branches of a species tree generated by a birth-and-death process of sufficient complexity. Cs\H{u}r\"os and Mikl\'os have recently developed an efficient algorithm for calculating the probability of observing a given phylogenetic profile as a function of branch-wise parameters of duplication, gain and loss along a species tree \cite{Csuros:2009zr}. Their model assumes that gene families evolve according to a linear birth-and-death process along the branches of the species tree. Each branch is characterized by a duplication rate, a gain rate, and a loss rate. A gene family evolves along the tree from the root toward the leaves according to the birth-and-death process. At internal nodes of the tree, families are instantaneously copied to evolve independently along descendant branches. Transient distributions of the linear version of processes presented in figure \ref{BD} give the expected change in the number of vertically inherited genes ( ``inparalogs'') and recently acquired ones (``xenologs'')\cite{csuros_arxiv}. \footnote{Leading up to the work of Cs\H{u}r\"os and Mikl\'os, other groups had also developed likelihood-based methodologies. These either only considered duplication and loss \cite{Hahn:2005uq} or relied on heuristic restrictions on maximal ancestral family size for computational tractability\cite{Spencer:2006kx,Iwasaki:2007fk}.}.

Using the above approach, it is possible to search for the branch-wise duplication, gain and loss rates that maximize the likelihood given a set of observed profiles (derived from complete genome sequences) and a species phylogeny. Conceptually, this is no different than searching for branch-wise substitution rates that maximize the likelihood given a set of homologous sites (see for instance Chapter 16 of \cite{Felsenstein:2004dq}). Columns of an alignment in the former case correspond to the phylogenetic profile of an individual gene family in the later. In table \ref{COUNT_TABLE} we present results obtained in this manner using COUNT \cite{Csuros:2010hc}, a software that provides an implementation of this calculation. The results in table \ref{COUNT_TABLE} lend further support to both the observation that birth-and-death rates are similar across the tree of life (although here we have only considered prokaryotes) and the pattern of death (loss) rates being on average significantly larger than birth (duplication and gain) rates. Similar to what was observed for 28 archaeal genomes \cite{Csuros:2009zr}, duplications are inferred to account for the majority of birth events. 

\begin{table}
\label{COUNT_TABLE}
\caption{ Relative rates of duplication, gain and loss for prokaryotic phyla obtained by maximum likelihood using COUNT \cite{Csuros:2010hc}. Rooted reference trees were obtained from concatenates of universal and near universal genes and phylogenetic profiles extracted from version 4 of the HOGENOM database \cite{Penel:2009ly}. Relative rates correspond to the ratio of the average of the branch-wise rates (of duplication, gain and loss) to the average branch-wise sum of the three rates. }
\begin{tabular}{l|c|c|c|c}
\textbf{phylum name} & \textbf{loss} & \textbf{duplication} & \textbf{gain} & \textbf{\# of genomes} \\
\hline\hline
Actinobacteria & 0.75 & 0.23 & 0.010 & 31 \\ 
 \hline
Alphaproteobacteria & 0.85 & 0.13 & 0.008 & 47 \\ 
 \hline
Bacillales & 0.52 & 0.42 & 0.048 & 16 \\ 
 \hline
Bacteroidetes/chlorobi & 0.59 & 0.38 & 0.024 & 10 \\ 
 \hline
Betaproteobacteria & 0.63 & 0.32 & 0.037 & 32 \\ 
 \hline
Chlamydiae/Verrucomicrobia & 0.70 & 0.24 & 0.043 & 7 \\ 
 \hline
Clostridia & 0.57 & 0.37 & 0.055 & 11 \\ 
 \hline
Cyanobacteria & 0.68 & 0.28 & 0.027 & 14 \\ 
 \hline
Deltaproteobacteria & 0.64 & 0.33 & 0.024 & 13 \\ 
 \hline
Epsilonproteobacteria & 0.54 & 0.29 & 0.158 & 7 \\ 
 \hline
Gammaproteobacteria & 0.88 & 0.10 & 0.009 & 70 \\ 
 \hline
Lactobacillales & 0.66 & 0.29 & 0.036 & 21 \\ 
 \hline
Mollicutes & 0.49 & 0.47 & 0.023 & 14 \\ 
 \hline
Spirochaetes & 0.79 & 0.19 & 0.014 & 7 \\ 
 \hline \hline
Crenarchaeota & 0.69 & 0.28 & 0.018 & 11 \\ 
 \hline
Euryarchaeota & 0.66 & 0.31 & 0.016 & 25 \\ 
 \hline \hline
\end{tabular}
\end{table}

\section{ The ubiquity of phylogenetic discord and the joint reconstruction of pattern and process}
In order to extract as much information as possible, we must step beyond phylogenetic profiles and consider in more detail the phylogenetic information contained in the sequences of homologous gene families. This can be done by using some model of sequence evolution to infer a gene phylogeny from the multiple sequence alignment (MSA) of the family. Because gene families evolve through not only the genome level process of speciation, but also the gene level processes of origination, duplication, transfer and loss described above, the phylogenies of individual families constructed in this manner will reflect intricate individual genic histories. Differences in the histories of individual families will inevitably lead to phylogenetic discord among gene families. The amount of phylogenetic conflict will reflect the extent of horizontal gene transfer among genomes, consequently the profusion of phylogenetic discord that we observe among prokaryotes (see below) is interpreted as reflecting large rates of transfer. 

Independent of the degree of HGT, however, the existence of gene level processes of birth and death make it necessary to extend the implicit model behind the tree of species. This extension consists of taking into consideration the processes of gene origination, birth and death described above. The classic concept of the species tree implicitly assumes that all genes evolve along a strictly shared track - the branches of the species tree. The presence of duplications, transfers and losses obliges us to replace this model by a tree, the branches of which can be best visualized as tubes -- tubes within which genes may duplicate and be lost, and among which they can be transferred. This tree of genomes is a straightforward extension of the classic tree of species with its branches characterized by rates of duplication, transfer and loss. 

For this tree of genomes to be useful, however, methods based on statistical models capable of considering data from complete genome sequences and inferring such a tree need to be developed. Below we describe recent progress in the construction of tractable models of genome evolution that are full, probabilistic models of all variables, in particular in our case of branch-wise duplication, transfer and loss rates and the species tree topology. 

\subsection{Phylogenetic discord among homologous gene families}
\label{phydiscord}
\begin{figure}
\label{discord}
\centerline{\includegraphics[width=1.\textwidth]{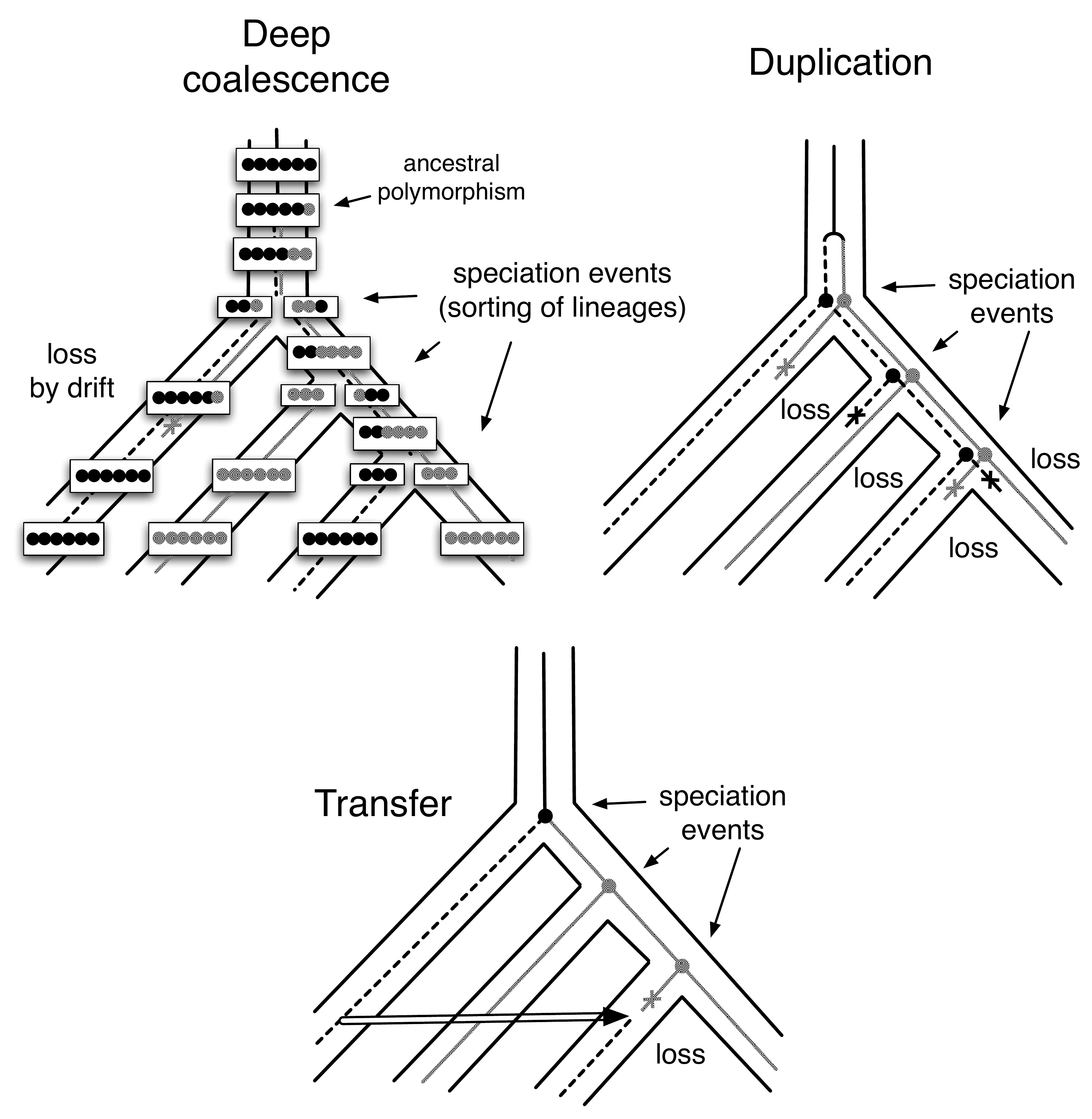}}
\caption{
\textbf{Evolutionary processes behind phylogenetic discord}
Phylogenetic incongruences can be the result of three major evolutionary processes \cite{Galtier:2008zr}: i) deep coalescence resulting from incomplete lineage sorting (see previous chapter); ii) hidden paralogy (resulting from duplication and differential loss); and iii) horizontal gene transfer (HGT). Incomplete lineage sorting occurs when an ancestral species undergoes two speciation events in rapid succesion. If, for a given gene, the ancestral polymorphism has not been fully resolved into two monophyletic lineages at the time of the second speciation, with a probability determined by the effective population size, the gene tree will differ from the species tree. A potential source of incongruence relevant over wider phylogenetic scales is hidden paralogy. If a gene family contains paralogous copies (genes that are related by a duplication event, e.g. the dashed and grey lines above), the gene phylogeny will partly reflect the duplication history of the gene that is independent of species divergence history. The third process is HGT. If genetic exchanges occur between species, then the phylogeny of individual genes will be influenced by the number and nature of transfers they have undergone. In the above figure we illustrate how a particular gene tree topology can be explained by each process. Depending on the parameters (duplication, transfer and loss rates and effective population size) describing the branches of the species tree the three different scenarios will have different probabilities. 
}
\end{figure}

Apparent phylogenetic conflict can result from different processes. First of all, inferred gene tree topologies can be different from the species tree, and hence each other, in the absence of any biological processes due to reconstruction errors. Such errors can result from stochastic differences caused by e.g.\ insufficient sequence length, and more problematically, from systematic reconstruction artifacts, due to departures from model assumptions \cite{Jeffroy:2006ve}. More informatively, phylogenetic discord can result from three important biological processes (summarized in figure \ref{discord}): lineage sorting, HGT and hidden paralogy. 

Galtier and Daubin \cite{Galtier:2008zr} analysed the level of phylogenetic conflict between genes in several datasets extracted from the from the HOGENOM \cite{Penel:2009ly} database. Their aim was to ascertain the relative contribution of HGT to the amount of phylogenetic discord by comparing metazoan datasets (where HGT can be assumed to be rare) to prokaryotic ones. Their results were consistent with expectations as the level of discord measured for metazoan sequences was smaller than for any of the bacterial datasets considered. Interestingly, however, the differences in the amount of discord among the bacterial datasets was also measured to be large (see table 1 of \cite{Galtier:2008zr}). These large differences in the amount of discord, presumably caused by differences in rates of transfer, stand in stark to the broadly similar rates of gene birth and death implied by the similarity of the gene family size distributions. 

A further finding of the study of Galtier and Daubin was that even in the case of actinobacteria (the prokaryotic dataset with the highest degree of self-conflict), more than 75 per cent of the genes did not significantly reject the consensus tree. While it is clear that including more and more species would cause this particular measure to converge to a much smaller value, a series of more careful studies have demonstrated that there exists a strong signal of vertical inheritance in prokaryotic genomes despite persistent HGT\cite{Daubin:2003vn,Ochman:2005ys,Beiko:2005fk,Puigbo:2009fk} (see also the next Chapter by Koonin et al.).

\subsection{Reconciling phylogenic discord}

The detection and measurement of phylogenetic discord among a group of phylogenetic trees can be accomplished relatively easily, for instance, by using some measure of distance between trees (see Chapter 30 of \cite{Felsenstein:2004dq} for an introduction on distance measures). A different and harder problem consists of constructing a \emph{reconciliation} between two trees, i.e., of proposing a set of evolutionary events (such as speciations, duplications, transfers and losses) that correspond to an evolutionary scenario where one of the trees (the gene tree) has resulted from evolution along the other tree (the species tree). In figure \ref{discord} we present three different reconciliations involving different sets of events for the same gene tree. The set of events considered in the context of the reconciliation problem has, until recently, been limited to speciation, duplication and loss events and models involving lineage sorting discussed in the previous chapter (and respectively Chapter 29 and 25 of \cite{Felsenstein:2004dq}). Goodman \cite{Goodman:1979qo} was the first to describe an algorithm to find the reconciliation that minimizes the number of duplication and loss events followed more recently by several others ( see \cite{Felsenstein:2004dq} for citations). If transfers are also considered, the problem of reconciliation becomes difficult from a combinatorial perspective for two reasons: i) the difficulty of restricting the set of events to ones which respect the partial order of evolution imposed by speciation events on the species tree\footnote{This corresponds to forbidding the transfer of genes from a species (branch of the species tree) to species from which it has descended (ancestral branches of the species tree), i.e., forbidding transfers that ``go backwards in time''. } \cite{Hallett2004ac}; ii) if transfer events are considered where the acquisition of a homologous copy implies the loss of extent copy, the problem of identifying the minimum number of such events can easily be shown to correspond to the problem of finding the shortest path between two trees using subtree prune and regraft (SPR) operations that is know to be NP-complete (see Chapter of 4 and 30 of \cite{Felsenstein:2004dq}).

The latter process of replacement of genes by HGT is biologically motivated by the elevated probability of functional redundancy in the case of homologous genes \cite{Abby:2010ve}. Such replacement is particularly relevant in modeling genes that are present in a single copy in all or most genomes. A variety of approaches have been put forward to solve the problem of tree reconciliation for the case when the replacement of genes is relevant \cite{Nakhleh:2005lr,Beiko:2006uq,Abby:2010ve}. These approaches offer heuristic algorithms to find approximate solutions to the SPR, and the closely related MAF (maximum agreement forest) problems efficiently. However, they are all limited to single label trees, i.e., trees for families that do not have multiple members in any of the genomes considered.

The former problem of considering only transfers that respect the partial time order implied by the species tree can be resolved by fully specifying the time order of speciation events. As shown by Tofigh et al.\ \cite{tofigh_2009}, and described below, this allows the construction of a dynamic programming algorithm that is able to efficiently traverse all possible reconciliations allowing the calculation of the sum of the probabilities of all reconciliations given a tree, the most parsimonious reconciliation \cite{doyon_2010,David:2011zr} or the reconciliation with the highest likelihood. 

\begin{figure}
\centerline{\includegraphics[width=0.7\textwidth]{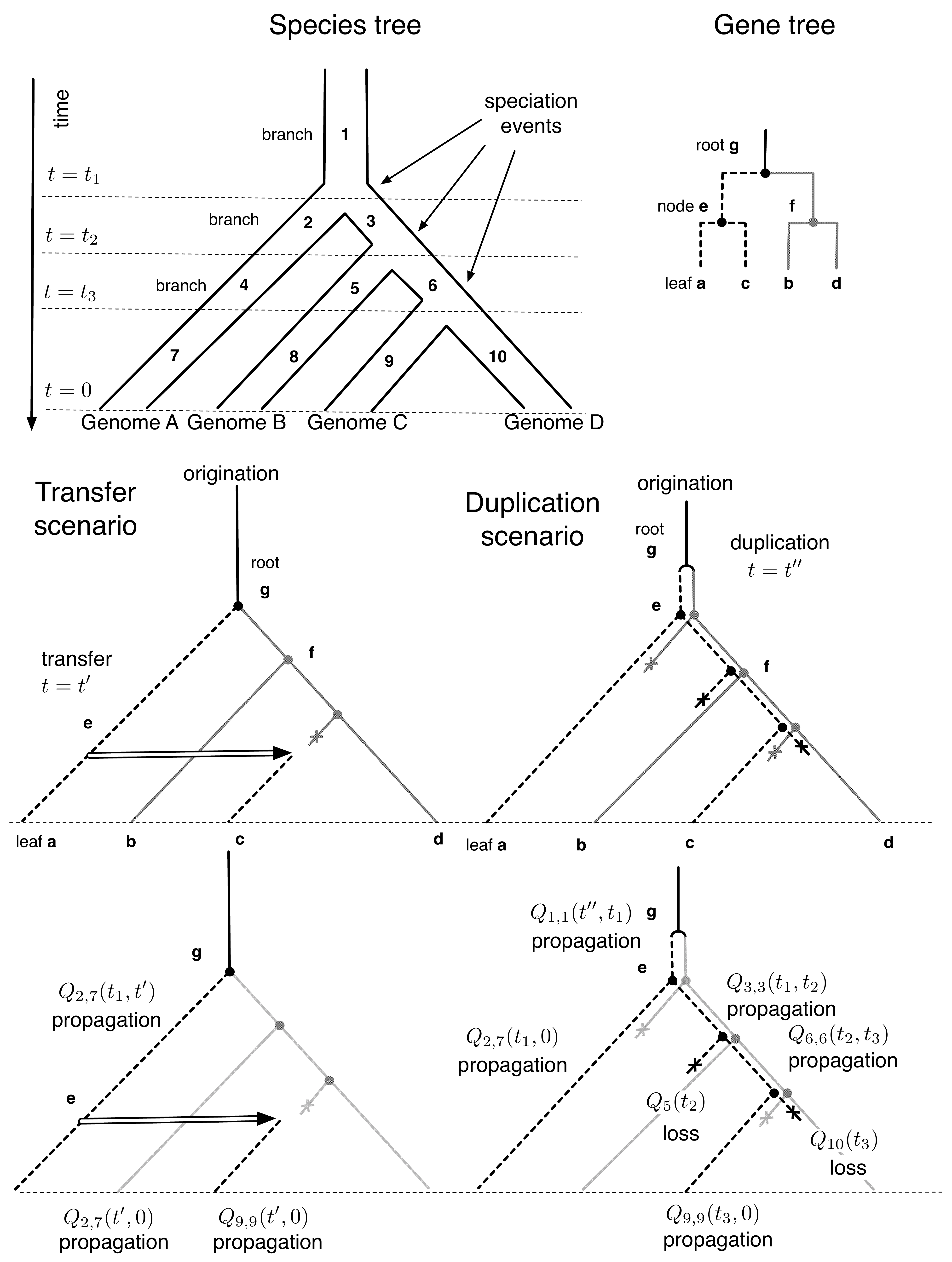}}
\caption{
\textbf{Probabilistic DTL model.} If we consider gene trees to be generated by a linear birth-and-death process $\mathcal{M}_\mathrm{BD}$ taking place on a tree $S'$ with the order of speciation events fully specified, we can express the probability of a gene tree topology $G$ given a reconciliation. Specifying the order of speciation events corresponds to constructing \emph{time slices}, which decompose the branches of the species tree into pieces yielding the tree $S'$. For example, the branch leading to Genome A is decomposed into three branches labeled 2,4,7 (for a formal definition see \cite{tofigh_2009}). Transfers are only possible between branches in the same time slice, e.g.\ between 7 and 9, but not 4 and 9. A reconciliation consists of mapping the branches and nodes of $G$ to the branches of nodes of $S'$. For a given gene tree there are many possible reconciliations. For $G$ we can construct i) a transfer scenario where node $g$ of $G$ is a speciation at the root of $S'$, $e$ is a transfer from 4 to 9,$f$ is a speciation at the end of 3, and the branch below $f$ traverses the speciation at the end of 6 implying at least one loss; and also ii) a duplication scenario, where $e$ maps to the root, $g$ is a duplication above it, the position of $f$ is unchanged, but at least four losses have occurred. The probability of extinction $Q_e(t)$ and the propagator $Q_{ef}(t,t')$ can be used to construct the probability of a given reconciliation as shown for the black subtree of $G$. Because the probability of a reconciliation can be hierarchically decomposed into the product of the probabilities of the reconstructions of the subtrees of $G$, a dynamic programming algorithm can be derived that is able to calculate the sum or maximum of the probability over all reconciliations.
}
\label{Qef}
\end{figure}

\subsection{The probability of a gene tree given a species tree and rates of Duplication, Transfer and Loss}

Tofigh et al.\ consider the forest of gene trees to be generated by a common birth-and-death process taking place on a shared species (or genome) tree. They derive the probability $p ( G | S' ,\mathcal{M _\mathrm{BD}},r)$ of a gene tree topology $G$ given a reconciliation $r$, where $\mathcal{M}_\mathrm{BD}$ is a birth-and-death process taking place on $S'$, a species tree for which the order of speciation events are fully specified. Provided the process $\mathcal{M}_\mathrm{BD}$ is linear, the probability of gene tree topology $G$ can be expressed given a reconciliation $r$ that maps branches and nodes of $G$ to $S'$ using events considered in $\mathcal{M}_\mathrm{BD}$. 

This calculation requires two functions: i) the probability of extinction $Q_e(t)$, i.e., the probability of a gene observed on branch $e$ at time $t$ evolving such that it is not observed in any extant genome (at time $t=0$); ii) the propagator $Q_{ef}(t,t')$ which gives the probability of a gene observed on branch $e$ at time $t$ evolving such that it has a descendent present on branch $f$ at time $t'$, further more any descendants of the gene observed at the leaves (at time $t=0$) of $S'$ descend from this copy. These functions can be obtained numerically from systems of differential equations found in \cite{tofigh_2009}. 

As illustrated in figure \ref{Qef}, the same gene tree can be reconciled in different ways with the species tree. The probability of extinction $Q_e(t)$ and the propagator $Q_{ef}(t,t')$ together with rates of origination, duplication and transfer can be used to calculate the probability of a gene tree topology for an arbitrary reconciliation\footnote{Here we present an example with a rooted gene tree, however, the position of the root can be considered to be part of the reconciliation without changing the complexity of the dynamic programing algorithm. }. For this probability to be useful, however we must be able to either sum over all reconciliations
\begin{equation}
\label{sum_p}
p ( G| S' ,\mathcal{M}_\mathrm{BD}) = \sum_{ r \in \Omega} p ( G | S' ,\mathcal{M _\mathrm{BD}},r)
\end{equation}
to obtain the probability of $G$ given $S'$ and $\mathcal{M}_\mathrm{BD}$, or alternatively be able to find the most likely reconciliation allowing the calculation of:
\begin{equation}
\label{max_p}
p_{max} ( G| S' ,\mathcal{M}_\mathrm{BD}) = \mmax_r p ( G | S' ,\mathcal{M _\mathrm{BD}},r).
\end{equation}

The probability of a reconciliation can be hierarchically decomposed into the product of the probabilities of the reconstructions of the subtrees of $G$. This allows the construction of a dynamic programing algorithm that can efficiently sum or take the maximum over reconciliations, allowing the calculation of both equation \ref{sum_p} and \ref{max_p}. Furthermore, the same dynamic programing scheme can be used to calculate the most parsimonious reconciliation given costs of the possible events with reduced complexity \cite{doyon_2010}.

\begin{figure}
\centerline{\includegraphics[width=1.\textwidth]{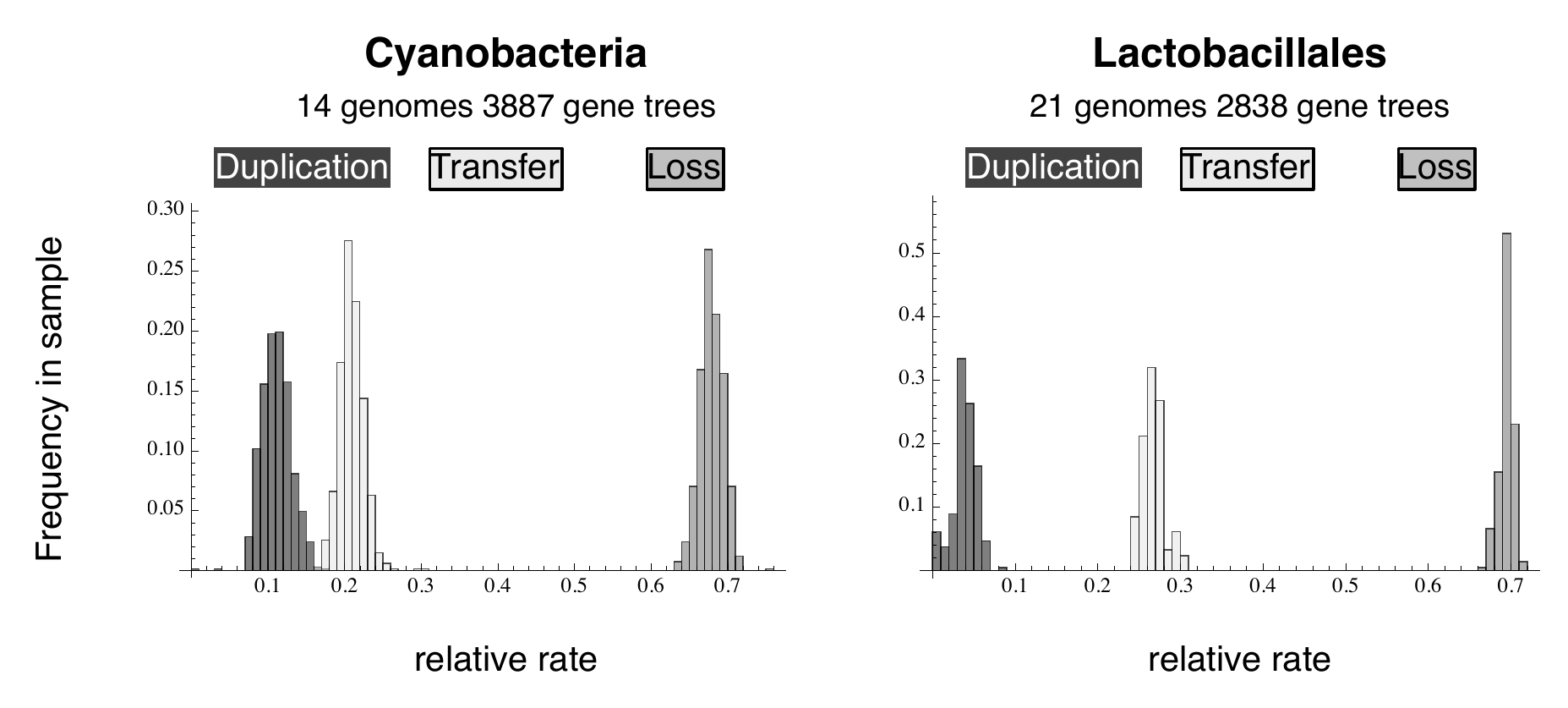}}
\caption{
\textbf{Relative rates of duplication, transfer and loss for two prokaryotic phyla.} The results were obtained by maximum likelihood using reference trees inferred from concatenated alignments of universal and near universal genes and all homologous gene families with trees available in version 4 of the HOGENOM database \cite{Penel:2009ly}. These results show that while the ratio of birth to death is practically identical, taking into consideration phylogenetic information from gene trees, the majority of birth events are inferred to have resulted from transfer and not duplication, in contrast to results obtained from phylogenetic profiles (see table \ref{COUNT_TABLE}). The histograms correspond to results obtained for $1000$ jackknife samples of $20\%$ all trees (see Chapter 20 of \cite{Felsenstein:2004dq} for a discussion of resampling). The calculation was implemented using results from \cite{tofigh_2009} and \cite{doyon_2010}. We kept the species tree topology fixed and maximized equation \ref{DTL_eq} over the space of possible orders in time of speciations and uniform rate parameters. We assumed each branch of $S'$ to have branch lengths compatible with the time order of speciations with all time slices being of equal width and inferred global rates of duplication, transfer and loss.
}
\label{DTL_plot}
\end{figure}

\subsection{Hierarchical probabilistic models of duplication, transfer and loss}

Using the above dynamic programing algorithm it is possible to calculate the likelihood of a species tree topology $S'$ and the parameters describing $\mathcal{M}_\mathrm{BD}$ , i.e., rates of duplication, transfer and loss on its branches, given a forest of gene trees obtained from homologous gene families:
\begin{eqnarray}
\label{DTL_eq}
\mathrm{L} ( S' , \mathcal{M}_\mathrm{BD}| \{G_f\} ) = \prod_{f \in \mathrm{ families }} p ( G_f| S' ,\mathcal{M}) \\
 \mathrm{where} \quad G_f= \argmax_G \left\{ \mathrm{L} ( G | \mathrm{\ MSA\ of\ } f\ ) \right\}, \nonumber 
\end{eqnarray}
and the product goes over the set of most likely gene trees $\{G_f\}$ encoding the sequence information in families of homologous genes composing a set of genomes. This expression can be thought of as being similar to the classic likelihood of a gene tree topology $G$ and some model of sequence evolution $\mathcal{M}_\mathrm{seq.}$ with parameters such as branch-wise substitution rates, given a multiple sequence alignment:
\begin{equation}
\label{seq_eq}
\mathrm{L} (G,\mathcal{M}_\mathrm{seq.} | \mathrm{\ MSA} ) = \prod_{i \in \mathrm{ sites }} p ( \mathrm{column\ } i \mathrm{\ of\ MSA } | G,\mathcal{M}_\mathrm{seq.}),
\end{equation}
where in this case the product goes over columns of homologous sites composing a multiple sequence alignment. In figure \ref{DTL_plot}, we present results obtained using such an approach, where we have kept the species tree topology fixed and maximized the likelihood given by equation \ref{DTL_eq} over the space of possible orders in time of speciations and uniform rate parameters. We can see that the inferred ratio of birth to death is in good agreement with that obtained from phylogenetic profiles (see table \ref{COUNT_TABLE}). In contrast, taking into consideration additional information from the sequences of the proteins in homologous families in the form of gene tree topologies, we infer for both phyla considered the majority of birth events to be the result of transfer. 

This scheme has two shortcomings. First, instead of complete sequence information only the most likely gene tree toplogies are considered. Second global information on how likely different gene tree toplogies are given $S'$ and $\mathcal{M}_\mathrm{BD}$ is not considered. Both of these shortcomings can be adressed by combining equations \ref{DTL_eq} and \ref{seq_eq} in a hierarchical likelihood framework. Using such a framework allows us to use global information on the species phylogeny and the birth-and-death process together with sequence information from each family to improve gene trees, while at the same time, inferring the species phylogeny and the parameters of birth-and-death process. Such a hierarchical framework was first suggested by Maddison \cite{Maddison:1997ly} and has recently been implemented using a duplication and loss model (excluding transfer) \cite{Akerborg:2009fk} and models of transfer (excluding duplication and loss) \cite{Suchard:2005nx,Bloomquist:2010oq}. The dynamic programming approach presents the first opportunity to construct a hierarchical model that considers all three processes. 
That is we can express the likelihood of $S'$, $\{G_f\}$ and $\mathcal{M}_\mathrm{BD}$ given a set of homologous gene families as 
\begin{equation}
\mathrm{L} ( S' , \{G_f\},\mathcal{M}_\mathrm{BD}| \mathrm{\ families\ }) = \prod_{f \in \mathrm{ families }} p (G_f | S' ,\mathcal{M}_\mathrm{BD} ) \times \mathrm{L} ( G_f | \mathrm{\ MSA\ of\ } f\ ).
\end{equation}
It is important to note that this hierarchical likelihood function is amicable to parallel computation, because the $p (G_f | S' ,\mathcal{M}_\mathrm{BD} ) \times \mathrm{L} ( G_f | \mathrm{\ MSA\ of\ } f\ ) $ terms can be computed independently, by client nodes. It is possible to implement an efficient optimization scheme consisting of a hierarchical optimization loop, wherein clients optimize the $G_f$-s using the independent terms in the hierarchical likelihood product while keeping $S$ and $\mathcal{M}_\mathrm{BD}$ fixed until conditionally optimal $G_f$-s are attained using which $S$ and $\mathcal{M}_\mathrm{BD}$ can be optimized.

\section{Conclusion}
In conclusion, the distribution of homologous gene family sizes in the genomes of the Eukaryota, Archaea and Bacteria show astonishingly similar shapes. These distributions are best described by models of gene family size evolution where the loss rates of individual genes are larger than their duplication rate, but new families are continually supplied to the genome by a process of origination that in general includes both transfer and the generation of new gene families. This picture is supported by analysis of phylogenetic profiles using maximum likelihood. Taking into consideration additional information from the sequences of the proteins in homologous families in the form of gene tree topologies, the inferred ratio of birth to death is found to be in good agreement with that obtained from phylogenetic profiles, however, in prokaryotes the majority of birth events is inferred to be the result of transfer. 

It has not been demonstrated to date that a single tree can adequately describe the evolution of entire genomes across the diversity of Life and certainly no such tree has been inferred. However, recent advances in the construction and implementation of hierarchical probabilistic models of duplication, transfer and loss presented here have the potential to allow us undertake this project, to infer genome trees based on sequence information from complete genomes. While currently this task is computationally daunting, the use of parallel computing and recent advances in algorithms present the promise of making this feasible in the foreseeable future. 

From a biological perspective, birth-and-death models of gene family size evolution are essentially neutral models of evolution. They ignore completely the individuality of gene families and any potential selective forces that make some of them expendable and others indispensable. The fact that they accurately reproduce the observed family size distributions nonetheless, suggests that genome evolution, at least on this coarse scale of observation, might be in large part the result of a stochastic process, which is only modulated by selection\cite{Koonin:2002rt,Lynch:2007bh}. Even so, as soon as we are able to better reconstruct the pattern and process of duplication, transfer and loss, we can expect to be able to observe more and more of this modulation by selection. And by proxy, start to learn more about the biology of genome evolution over large time scales, to better understand the population genetic, biochemical and ecological constraints and opportunities that govern the evolution of genomes in general and the transfers of genes in particular. This will require integrating information reconstructed from ancestral genomes and DTL events with system level models of phenotype such as metabolic networks \cite{Wagner:2009tw,Boussau:2010cr}. 

\bibliography{DTL}

\section{Exercises}
\begin{enumerate} 

\item Using log-log axis on the range $[ 0.1,10^6 ]$ plot the following functions: $e^{-x}$,$e^{-x/10}$,\\$e^{-x/100}$,$e^{- x /1000}$,$x^{-1}$,$x^{-3}$,$x^{-9}$ and observe how power-law like tails decay much slower than any exponential function.
 
\item Using both the COG and the HOGENOM databases\footnote{ You can find both these databases online at http://www.ncbi.nlm.nih.gov/COG/ and http://pbil.univ-lyon1.fr/databases/HOGENOM .} construct the histogram in figure \ref{BD_plots} of the frequency of homologous gene family sizes in the Human genome, i.e., the fraction $f_n$ of times you see a family of size $n$ among all homologous gene families in the Human genome. 

\item Using the result that the stationary distribution $p_n$ of family sizes is reached exponentially fast, and assuming this occurs according to the relationship $|p_n(t)-p_n| \propto \mathrm{e}^{-(\delta+\lambda)t}$, considering the rates of duplication and loss from table 8.1 of \cite{Lynch:2007bh} estimate the amount of time (in units of percentage of divergence at silent sites) that the distribution of family sizes needs to reach the stationarity distribution following a perturbation. Is this number large or small? Which organisms can be described by such divergence in comparison to the human genome?

\item Using the form of the transient distribution for the linear duplication-loss process given in table. 1. of \cite{csuros_arxiv} express the duplication rate $\lambda$ and the loss rate $\delta$ using the fraction of families with 0 genes and the mean number of genes in a family. 
 
\item Write down the differential equation giving $p(t)$ the probability of families with size $n$ at time $t$, using only the probabilities of $p_{n-1}(t)$, $p_{n+1}(t)$ and the rates of duplication $\delta_n=\delta n$ and loss $\lambda_n=\lambda n$ (note the case $p_0(t)$ needs to be treated differently, solution can be found in \cite{Reed:2004bh}). 

\item Using the transient distribution of the duplication-loss process from table 1 and the results of Lemma 1 in \cite{csuros_arxiv}, and assuming the species tree to be ((A:$y$,B:$y$):$x$,C:$x+y$) with branch lengths $x,y$ in arbitrary units of time, (see \cite{Felsenstein:2004dq} for a description of the Newick format), a duplication rate of $\delta$ ,a loss rate of $\lambda$ and assuming the probability of the number of genes in a family at the root of the tree to be given by a Poisson distribution with mean $n_0$, further limiting the number of genes at internal nodes to a maximum of $M$ genes, derive the probability of observing a profile $\{n_A,n_B,n_C\}$.
 
\item In what respect would including gain introduce significant new complications in the above calculations?

\item Considering only duplications and losses (excluding transfer) express $Q_{ef}(t,t')$ using transient distributions from \cite{csuros_arxiv} and the extinction probability $Q_e(t)$.

\end{enumerate} 

\end{document}